# Structural properties and stability of defected ZnSe/GaAs(001) interfaces

A. Stroppa and M. Peressi

INFM-DEMOCRITOS *National Simulation Center,*

*Trieste, Italy, and Dip. Fisica Teorica,*

*Univ. Trieste, I-34014 (Ts), Italy*

## Abstract

Accurate ab-initio pseudopotential calculations within density functional theory in the LDA approximation have been performed for structural properties and stability of ZnSe/GaAs(001) defected heterostructures. There is a strong experimental evidence that ZnSe/GaAs heterostructures with minimum stacking fault density are related to the presence of a substantial concentration of Ga vacancies at interface. In order to gain insights into the still unknown microscopic mechanism governing their formation and stability, we compared the relative stability of some simple selected interface configurations, chosen taking into account charge neutrality prescription and allowing the presence of Ga vacancy next to the interface. Remarkably, our results show that, under particular thermodynamic conditions, some interfaces with vacancies are favoured over undefected ones.

# 1 Introduction

II-VI/III-V heterostructures, whose prototype is ZnSe/GaAs, fabricated by molecular beam epitaxy (MBE) are important systems as they are the backbone of a certain number of electronic devices which are nowadays under development, mainly blue-green emitters [1] in optoelectronics, spin-transistors [2,3] and spin filters [4] in spintronics.

For all the proposed applications, is crucial to have a control of native defect density and, in general, of the structural quality of these interfaces. To develop high-quality heterostructures, much effort has been devoted in establishing the particular growth conditions for the reduction of the native stacking fault density (SF) [5] which plays an important role in device degradation.

It is widely accepted that the driving force which determines the SF density in II-VI/III-V interface depends upon the conditions in which the interface is fabricated. Only two procedures so far established were found to work for this purpose, yielding quantitatively similar defect densities (below $10^4$ cm$^{-2}$, thus providing a very high quality system), and qualitatively similar interface compositions and band alignments [6].

Apart from this experimental evidence, the microscopic mechanisms that control the native defect density in II-VI/III-V structures remain still controversial. What is noteworthy is that from an accurate characterization of the samples with minimum SF density there is a strong experimental evidence of the formation of a ternary (Zn,Ga)Se alloy of variable composition with a substantial concentration of cation vacancies (V) [6]. Because of vacancies, its average lattice parameter is smaller than the one of GaAs and ZnSe, and therefore this alloy is under tensile biaxial strain when epitaxially grown on GaAs substrates, accumulating a non negligible elastic energy. The driving mechanism for the formation of this defected alloy at ZnSe/GaAs interfaces is still unknown and the reason why it can be favourite compared with the simplest case of undefected unstrained interface with simply cation and/or anion mixing has to be investigated.

In this paper, we present therefore a comparative study based on ab-initio local density-functional pseudopotential approach of selected interface morphologies of ZnSe/GaAs (001) interfaces, also including vacancies, with the purpose of identifying some realistic configurations which can be related with the observed (Zn,Ga)Se interface compound formation. It is well assessed that abrupt polar interfaces are charged and energetically unstable with respect to the interdiffusion of atoms across the interface [7-12] and therefore it is mandatory to consider intermixed and reconstructed interfaces to satisfy the charge neutrality condition. We will consider different cases of neutral interfaces with vacancies, starting from the simplest one of vacancies confined just in the nominal interface plane. Our comparative study will address their *interface formation energy* and their structural properties.

This work is organized as follows: in the next section we describe the theoretical and computational approach; in Sect. III we discuss the selected interface morphologies; in Sect. IV,V we report our results; finally, in Sect. VI we present our conclusions.

## 2 Theoretical and computational method

Our calculations are performed within the density functional theory framework using the local density approximation for the exchange-correlation functional [13,14] with state-of-the-art first-principles pseudopotential self-consistent calculations [15]. The theoretical lattice constant of GaAs is 1.8 % smaller compared with the experimental lattice constant ($a^{GaAs}_{exp}$=0.565 nm [16]), whereas for ZnSe the lattice constant is 1.4 % smaller than the experimental one ($a^{ZnSe}_{exp}$=0.567 nm [16]), but the *relative* difference between the two bulks is well reproduced.

The interfaces are modeled by tetragonal supercells with periodic boundary conditions. The in-plane lattice constant is fixed to the theoretical lattice constant of GaAs which is the substrate.

The ionic degrees of freedom are fully taken into account by optimizing the atomic positions via a total-energy and atomic-force minimization [17].

We study the relative stability by considering the interface formation energy $E_{intf}^{Form}$ [18]:

$$2E_{intf}^{Form} = \frac{\left(E_{supercell} - \sum_{i=1}^{N_{species}} n^i \mu^i\right)}{mn} \qquad (1)$$

where $E_{supercell}$ is the calculated total energy of the supercell, $N_{species}$ is the number of the chemical species involved (which are 4 in the present case), $n^i$ the number of atoms of the species $i$ and $\mu^i$ is the corresponding chemical potential; ($m \times n$) is the reconstruction (the in-plane periodicity of the interface), so that $N = mn$ is the number of atoms per layer parallel to the interface and $E_{intf}^{Form}$ refers to a (1×1) interface area. In general $E_{intf}^{Form}$ refers to a *mean* value of the formation energy of the two possibly inequivalent interfaces present in each supercell. We calculated the bulk chemical potentials considering the elemental forms of Zn (hcp) [19], Se(trigonal) [19], Ga (orthorhombic) [19], As (trigonal) [19], GaAs and ZnSe (cubic) [16].

## 3  Interface structures

We start with the simplest cases with only one mixed atomic plane, including the possibility of vacancies, satisfying the charge neutrality. The composition profiles are:

*(i)* single-plane cation-mixed interface: . . . -Ga-As-[Ga$_x$ Zn$_y$ V$_{1-x-y}$ ]-Se-Zn-. . .

*(ii)* single-plane anion-mixed interface: . . . -As-Ga-[As$_w$ Se$_z$ V$_{1-w-z}$]-Zn-Se-. . .

where *x*, *y*, *w*, and *z* are the fractions of the respective atomic species (with x+y ≤1 and w+z ≤1) and V indicates vacancies. Considering the ionic charge associated to each atomic plane and handling with integer numbers for atoms of one or another species, we have for the single-plane cation and anion mixed interface the equations: 6X+4Y = 5N and 10W+12Z = 11N with X+Y ≤ N and W+Z≤ N, where N = $m \times n$ is, as defined above, the number of atoms per layer parallel to the nominal interface plane, and X = N$x$, Y = N$y$, W = N$w$ and Z = N$z$ are integer numbers.

The interface structures with no vacancies, i.e. with X+Y = N (or W+Z= N), correspond to the well known cases of 50%-50% single-plane cation (or anion) mixed interface already widely studied in the literature [12].

Considering the presence of vacancies, many other different morphologies compatible with the charge neutrality condition are possible. In order to restrict our case studies, we take advantage of the experimental evidence of the formation of the (Zn,Ga)Se compound with cation vacancies, without As, and therefore we neglect cases with anion mixing. We generalize instead the case of single-plane cation-mixed interface by considering the possibility of another mixed cation plane adjacent to pure Se planes as follows:

...-Ga-As-[Ga$_x$ Zn$_y$V$_{1-x-y}$ ]-Se-[Ga$_w$ Zn$_z$V$_{1-w-z}$ ]-Se - Zn- ...

The charge neutrality condition applied to this composition profile gives:

$6(X+W)+4(Y+Z) = 9N$. Beside the solution with $z=1$, $w=0$, which reduces to the single-plane cation-mixed case already discussed, there are non trivial solutions with vacancies already for $N=2$, which are (one is complementary to the other, since x can be interchanged with w and y with z):

($\beta$): $x = 0.5$, $w = 1$, $y = z = 0$: ...- Ga - As - Ga $_{1/2}$ - Se - Ga - Se - Zn- ...;

($\beta_1$): $x = 1$, $w = 0.5$, $y = z = 0$: ...- Ga - As- Ga- Se - Ga $_{1/2}$ - Se - Zn - ... .

Next, we select $N=6$, and out of all the possible solutions we consider the following ones with vacancies:

($\alpha$): $x = \frac{2}{3}$, $w = \frac{5}{6}$, $y = z = 0$: ... - Ga - As - Ga $_{2/3}$ - Se - Ga $_{5/6}$ - Se - Zn -...;

($\alpha_1$): $x = \frac{5}{6}$, $w = \frac{2}{3}$, $y = z = 0$: ... - Ga - As - Ga $_{5/6}$ - Se - Ga $_{2/3}$ - Se - Zn -...;

($\alpha_2$): $w = \frac{5}{6}$, $y = 1$, $x = z = 0$: ... - Ga - As - Zn - Se - Ga $_{5/6}$ - Se - Zn -...;

($\alpha_3$): $x = \frac{5}{6}$, $z = 1$, $y = w = 0$: ... - Ga - As - Ga $_{5/6}$ - Se - Zn - Se - Zn - ....

These cases can be classified into 2 groups: the former which contains interfaces with a 50% vacancy layer ($\beta, \beta_1$); the latter includes interfaces having planes with a different (and less than 50%) vacancy concentration ($\alpha_i$, i=1,2,3,4). We can further group the $\alpha_i$'s as $\alpha''$ and $\alpha'$ where $\alpha''$ indicate both $\alpha$ and $\alpha_1$ interfaces (which contain 2 vacant planes) and $\alpha'$ refers to both $\alpha_2$ and $\alpha_3$ (which contain one vacant plane). By inspection, it can be seen that $\alpha_2$ ($\alpha_3$) is obtained from $\alpha(\alpha_1)$ with the isovalent substitution $Ga_{2/3} \to Zn$. All the interfaces considered are characterized by the presence of Ga-Se bonds, which are not present in the bulk constituents. In general, at interfaces between heterovalent constituents there are "wrong" chemical bonds [21] with either more ("donor bond") or less ("acceptor bond") than two electrons. The charge neutrality rule can be equivalently expressed in terms of compensation of donor and acceptor bonds through reconstruction and atomic intermixing.

## 4 Structural properties

### 4.1 Interfaces with 50% vacancies layer

Both $\beta$ and $\beta_1$ supercells are characterized by a huge local concentration of cation vacancies, having a plane with 50% of vacancies. Because of that, atomic relaxations are sizeable in the interface region, involving in particular acceptor and donor bonds, and have a non negligible effect on the stability of the system. In both $\beta$ and $\beta_1$, symmetry allows relaxations mainly in the anion sublattice. These relaxations result in sizeable variations of the interatomic distances, as reported in Fig. 1.

In Fig. 2, we show the relaxed interplanar distances along the growth direction. In order to take into account the *buckling* of atomic planes (due to atoms of the same layer which move in a different away along [001] direction), we consider an average interplanar distance with an error determined by the maximum and minimum interlayer distance between adjacent atomic layers. However, for the systems considered here, the buckling is negligible.

Variations of the mean interplanar distance are larger in $\beta$ than in $\beta_1$ supercell. In the former the deviations also extend further from the vacant planes than in the latter. In the $\beta$ structure the $Ga_{0.5}$ and As atomic plane move in such a way as to reduce by 50% the interplanar $Ga_{0.5}$-As plane whereas the $Ga_{0.5}$-Se interplanar distance elongates by 21% with respect to the ideal unrelaxed value.

## 4.2 Interfaces with different vacancy concentration

The $\alpha_i$, i=1,...,4 interfaces have 6 atoms per layer and the supercell is made up by 20 atomic layers. In Fig. 3 we show the mean interplanar distances as they vary along the [001] direction. In this cases, the buckling of atomic planes is non negligible, in particular in the interface region: the spreading associated to the average interplanar distance (which is the amplitude of the buckling) decreases as the distance with respect to interface increases. The effect is visible also several atomic layers away from the interface region (long range relaxation effects) and much more in $\alpha''$ than in $\alpha'$. The pattern of distances is asymmetric with respect to a plane midway the two interfaces in the supercells, indicating that the two interfaces of each supercell are non equivalent even if the cation profile is symmetric. A careful analysis of the structural models justifies such asymmetry, since the two interfaces are not fully equivalent. Incidentally we note that a larger buckling of the interface planes is associated to large lattice distortions in the bulk slabs.

## 5 Thermodinamic stability of interfaces

We take care of the relative stability of the defected interfaces so far considered, using Eq. (1). Althought the precise values of the chemical potentials are unknown (they strongly depend on the growing process and on the local environment), one can set precise relationships and boundary conditions for their range of variation (or of their combinations) as long as the interface is in thermodynamic equilibrium. In particular: $\mu^{GaAs}_{bulk} = \mu^{Ga} + \mu^{As}$ and $\mu^{ZnSe}_{bulk} = \mu^{Zn} + \mu^{Se}$ (equilibrium between the interface and the bulks); $\mu^i \leq \mu^i_{bulk}$ (since when $\mu^i = \mu^i_{bulk}$ the gas phase condensates to form the elemental bulk phase); by defining the heat of formation of A-B compound $\Delta H^{AB} = \mu^{AB}_{bulk} - \mu^{A}_{bulk} - \mu^{B}_{bulk}$ (see Ref. [12]; a negative $\Delta H^{AB}$ means exothermic reaction), one gets the following relations: $\Delta H^{GaAs} + \mu^{As}_{bulk} \leq \mu^{As} \leq \mu^{As}_{bulk}$ and $\Delta H^{ZnSe} + \mu^{Se}_{bulk} \leq \mu^{Se} \leq \mu^{Se}_{bulk}$. So, we end up with the following range of variability for instance for $\mu^{Zn} - \mu^{Ga}$ and for $\mu^{Se}$ : $\mu^{Zn}_{bulk} - \mu^{Ga}_{bulk} + \Delta H^{ZnSe} < \mu^{Zn} - \mu^{Ga} < \mu^{Zn}_{bulk} - \mu^{Ga}_{bulk} - \Delta H^{GaAs}$ and $\Delta H^{ZnSe} + \mu^{Se}_{bulk} < \mu^{Se} < \mu^{Se}_{bulk}$. We will discuss the relative stability of the interfaces within these boundaries.

We can specify Eq. (1) as follows:

$$2E^{Form}_{intf} = \frac{E_{supercell} - N_{Se}\mu^{ZnSe}_{bulk} - N_{As}\mu^{GaAs}_{bulk} + (N_{Se} - N_{Zn})\mu^{Zn} + (N_{As} - N_{Ga})\mu^{Ga}}{mn} \quad (2)$$

Supercells containing the single mixed planes with no vacancies are stoichiometric ($N_{se}=N_{Zn}=N_{ZnSe}$, and $N_{Ga}=N_{As}=N_{GaAs}$), therefore their formation energy is independent on the chemical potentials. So Eq.(2) reduces to a constant value. The resulting interface formation energies for the undefected interface with single anion or cation mixed plane are nearly degenerate, within 10 meV per (1×1) interface unit cell, which can be estimated as our numerical uncertainty. The relaxations lower the interface formation energies by 0.060(0.050) eV per interface unit cell for the cation(anion)-mixed system.

The structures $\beta$ and $\beta_1$ have the same $N_i$ 's, differing only in the interface morphology. From Eq. (2) we get:

$$E_{form}^{intf,\beta(\beta I)} = C_{\beta(\beta 1)} + x + \frac{1}{2} y \quad (3)$$

where $x=\mu^{Zn} - \mu^{Ga}$ and $y=\mu^{Se}$, and $C_{\beta(\beta 1)}$ is a constant characteristic of the interface considered and independent on the lenght of the supercell used in the calculation. Incidentally, we note that this term change by only ~ 10 meV per (1×1) interface unit cell, going from c/a=4 to 6 for each β and $β_1$ system thus indicating that the size of our supercells is large enough to this aim. The structures α and $α_1$ ($α_2$ and $α_3$) have also the same $N_i$'s, and we get respectively:

$$E_{form}^{intf, α(α1)} = C_{α(α1)} + x - \frac{1}{2} y \qquad (4)$$

$$E_{form}^{intf, α2(α3)} = C_{α2(α3)} + \frac{1}{3} x - \frac{1}{6} y \qquad (5)$$

As can be seen from Eqs.(3-5), the interface energy $E_{intf}^{Form}$ is a linear function of x and y with coefficients determined by the particular stoichiometry. It is interesting to discuss the limiting cases of the lowest and uppermost values of $μ^{Se}$ ($μ^{Se} = μ^{Se}_{bulk}$, $ΔH^{ZnSe} + μ^{Se}_{bulk}$) (see Fig. 4). $E_{intf}^{Form}$ is function of the remaining chemical potential. Incidentally, the heat of formation for ZnSe is used to determine the lower limit of $μ^{Se}$.

It can be seen that the most stable structures remain the same ($β_1$, mixed-cation, mixed-anion) considering the lowest or uppermost values for $μ^{Se}$, although the value of $x=μ^{Zn}-μ^{Ga}$ when they are competing changes a little bit. Substantial changes occur instead considering the extreme cases for $x=μ^{Zn}-μ^{Ga}$. This indicates that the variation of $μ^{Se}$ has much less effect on the structure than a variation of $μ^{Zn}-μ^{Ga}$. It is interesting to note that in the high $μ^{Se}$ limit, the defected interfaces are favored over the mixed ones for almost the whole range of variation of x. On the contrary, the variation of $\mu^{Zn}$-$\mu^{Ga}$ has much stronger effect: in the limit of high $\mu^{Zn}$-$\mu^{Ga}$, all the defected interfaces turn out to be unstable over the mixed-ones, irrespectively on the variation of $\mu^{Se}$ ; in the limit of low $\mu^{Zn}$-$\mu^{Ga}$, almost all defected interfaces are favored over the mixed-ones, except $\alpha_2$. So a low $\mu^{Zn}$- $\mu^{Ga}$ growing condition should favour the formation of defected interfaces over the mixed ones.

We observe that $\beta$, $\beta_1$, $\alpha$, $\alpha_1$ (characterized by having no Zn-As bonds) have the same linear dependence on x and y so that we can compare easily their relative stability. Their relative order, from the most stable, is: $\beta_1$, $\alpha_1$, $\alpha$, $\beta$ . If we define an average number of "wrong bonds" (including the fictitious bonds defined with the vacancies) per anion in the supercell, we can empirically observe that, at least as far as $\beta_1$, $\alpha_1$, $\alpha$, $\beta$ are concerned, the most stable interface is the one with the lowest number of Se-Ga and As-V bonds and the highest number of Se-V bonds.

## 6 Conclusion

We have studied with accurate ab-initio calculations several selected models of interface configurations for the ZnSe/GaAs(001) heterojunctions, also with cation vacancies, but in any case within the conditions of charge neutrality. We have shown that in some particular thermodynamic conditions the formation of defected interfaces with cation vacancies is favoured over the more ideal, undefected, unstrained, single-plane anion- or cation-mixed interfaces. These results support recent experimental evidence of (Zn,Ga)Se defected compounds in high-quality (i.e. with low native stacking fault density) ZnSe/GaAs(001) heterojunctions.

# 7 Acknowledgments

Computational resources from CINECA (Bologna, Italy) have been used thanks to the ``Iniziativa Trasversale di Calcolo Parallelo'' of the Italian Institute for the Physics of Matter (Istituto Nazionale per la Fisica della Materia, INFM).

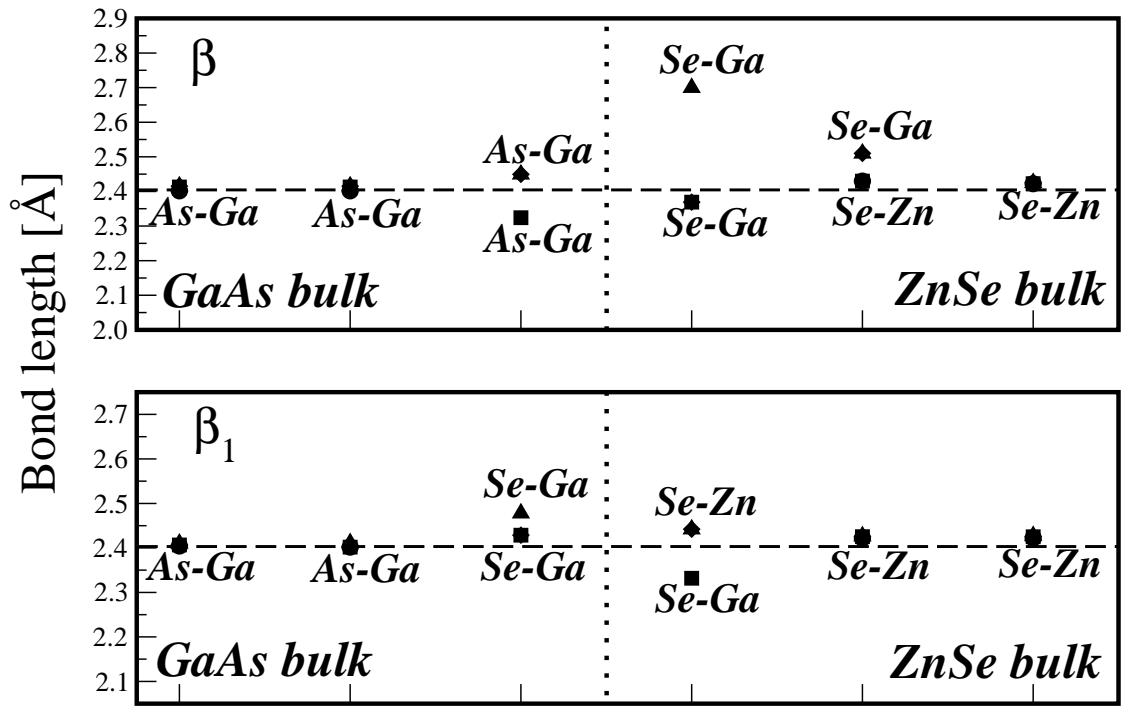

Figure 1: Bond lengths in β (top) and $β_1$ (bottom) supercells between atoms in adjacent planes parallel to the interface. Vertical dotted line denotes position of the plane with vacancies. Tick marks on x axis indicate anion planes. Dashed horizontal line refers to unrelaxed bulk-like bond-lenghts Ga-As or Zn-Se.

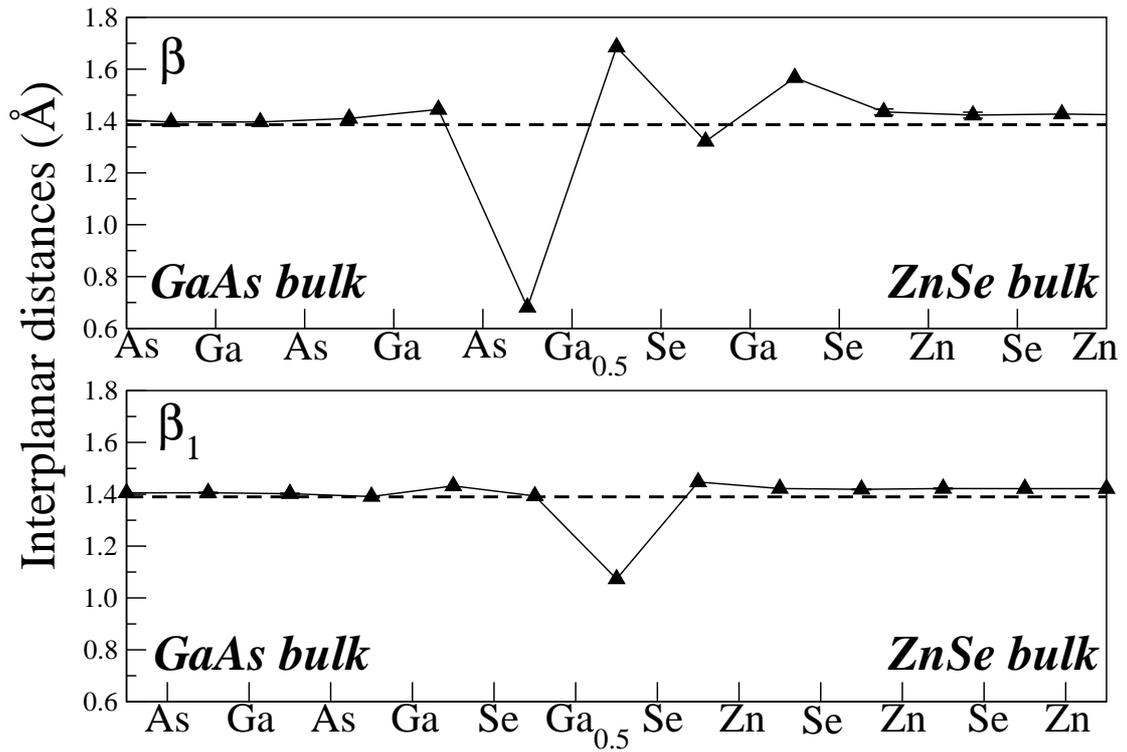

Figure 2: Interplanar distances in β ($β_1$) supercells as a function of the distance from the plane ($Ga_{0.5}$) containing Ga vacancies.

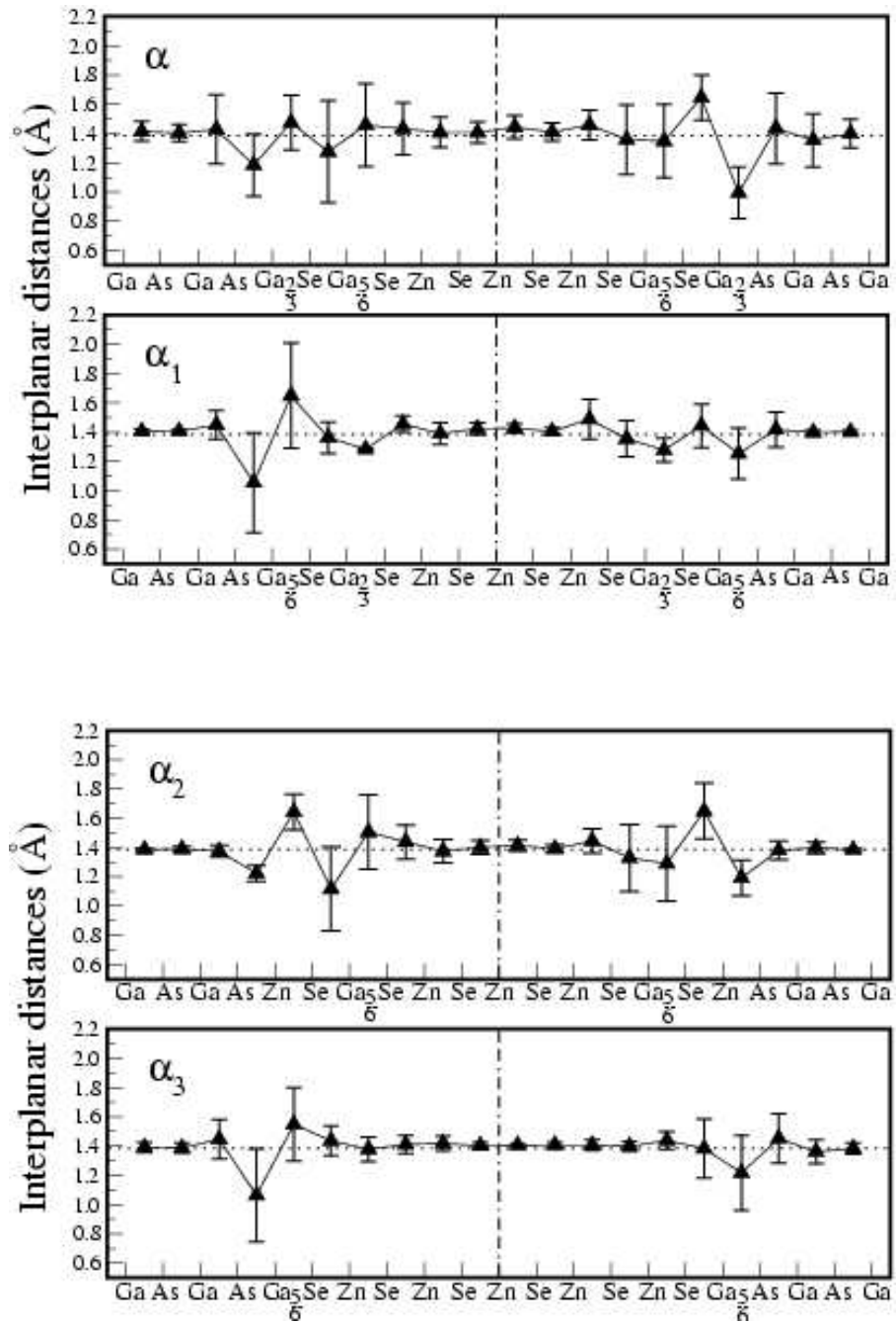

Figure 3: Interplanar distances for different structures considered ($\alpha_i$, i=1,4). The closed symbols and their associated error define the average and maximum (or minimum) distance between adjacent atomic layers, due to their possible buckling.

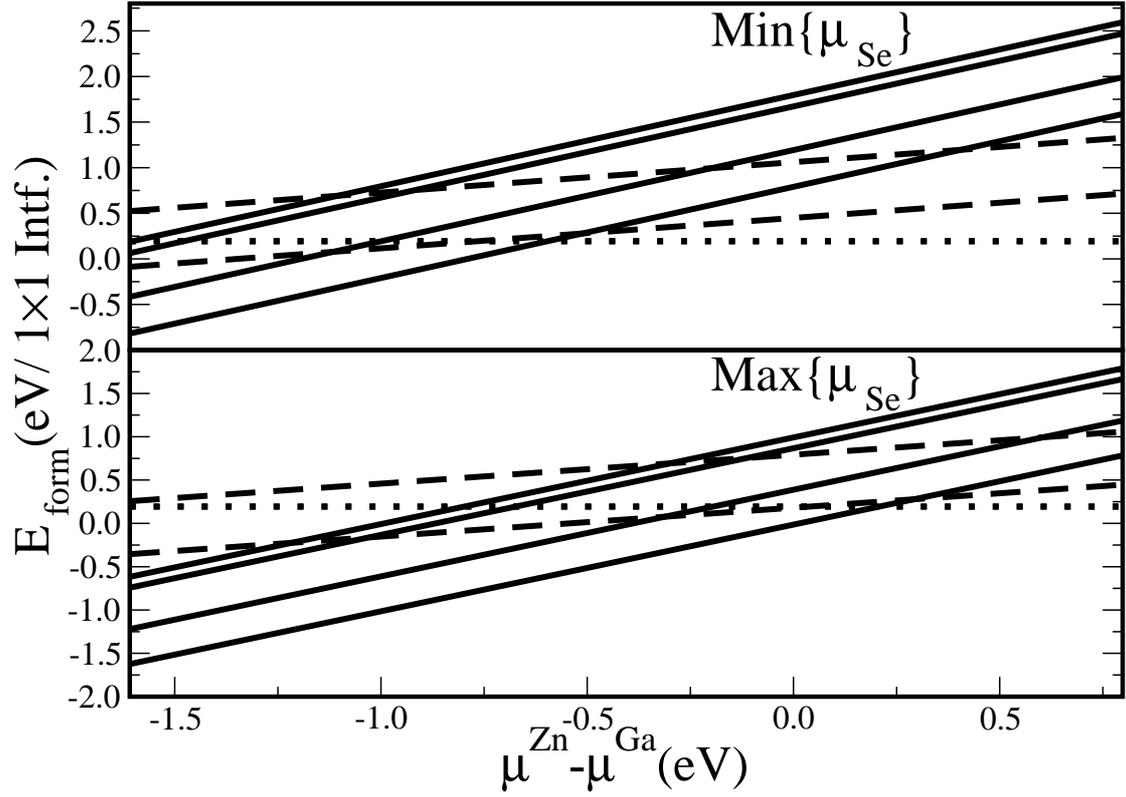

Figure 4: Formation energy [eV/(1×1) interface] of the mixed and defected interfaces as function of the difference of the Zn and Ga chemical potentials, for lower (upper) limit of $\mu^{Se}$. Solid lines are for $\beta$, $\alpha$, $\alpha_1$, $\beta_1$ and dashed lines for $\alpha_2$ and $\alpha_3$ respectively, in order of decreasing energy. Dotted lines refer to the mix-cation and mix-anion interfaces, which appear degenerate in the scale used here. $\mu^{Zn}-\mu^{Ga}$ has been rescaled with respect to $\mu^{Zn}_{bulk} - \mu^{Ga}_{bulk}$.

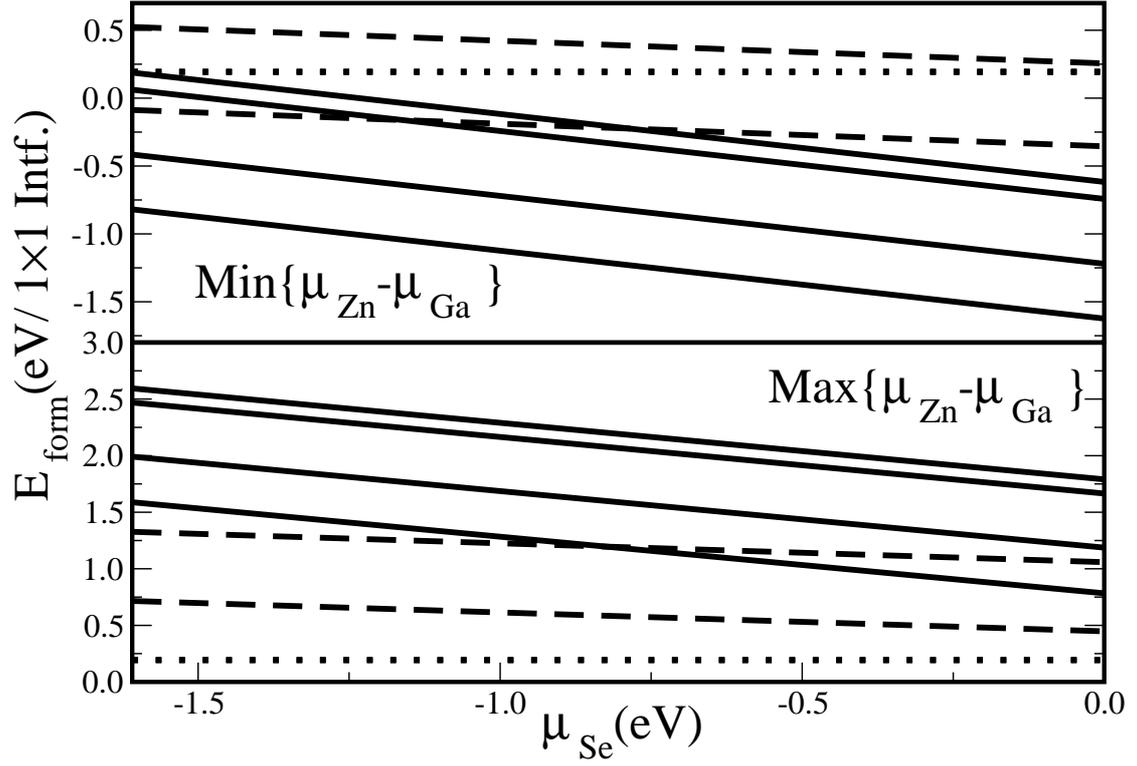

Figure 5: Formation energy [eV/(1×1) interface] of the mixed and defected interfaces as function of the Se chemical potentials for the lower (upper) limit of $\mu^{Zn}-\mu^{Ga}$. Solid lines are for $\beta$, $\alpha$, $\alpha_1$, $\beta_1$ and dashed lines for $\alpha_2$ and $\alpha_3$ respectively, in order of decreasing energy. Dotted lines refer to the mix-cation and mix-anion interfaces, which appear degenerate in the scale used here. $\mu^{Se}$ has been rescaled with respect to its own bulk value $\mu^{Se}_{bulk}$.